% ------------------------------------------------------------------
%    dirhak.tex    12 May 2005
% ------------------------------------------------------------------
\def\ptitle{Iterative solutions to the Dirac equation}
% ------------------------------------------------------------------------
% Change list publication -> preprint
% (1) Magnification : either magstep1 and 10pt fonts, or no mag and 12pt
% (2) Remove \parskip and set \baselineskip = 16pt before Abstract
% (3) Remove \np before (but not after) references
% (4) Add \input psfig.sty (or some such) at top
%     Add \psfig commands for each figure
% ------------------------------------------------------------------------
\input psfig.sty

\nopagenumbers
%\magnification=\magstep1
\hsize 6.0 true in 
\hoffset 0.25 true in 
% 6 in width with 1.25 in margins default = (6.5, 0)
\emergencystretch=0.6 in                 % TEXBook p 107 : allows h-space 
\vfuzz 0.4 in                            % page-length flexibility
\hfuzz  0.4 in                           % line-length flexibility
\vglue 0.1true in
\mathsurround=2pt                        % Default is 2pt
\topskip=24pt                            % Default is 10pt
\def\nl{\noindent}                       % New line after equations
\def\np{\hfil\vfil\break}                % New page
\def\ppl#1{{\leftskip=9cm\noindent #1\smallskip}} % Preprint line 
\def\title#1{\bigskip\noindent\bf #1 ~ \tr\smallskip} % Headings
% --------------------------------------------------------------------
%   PC Fonts (used only locally)
% --------------------------------------------------------------------
%\font\tr=TIMENRR                       % Times New Roman: see output
%\font\bf=TIMENRB                       % Redefinition
%\font\it=TIMENRRI                      % Redefinition       
%\font\trbig=TIMENRR scaled \magstep3   % For main Title
%\font\th=CMBXSL10                      % Theorems                    
%\font\tiny=CMBX8                       % Page numbers
% --------------------------------------------------------------------
%  generic unix fonts (lower case names)
% --------------------------------------------------------------------
\font\tr=cmr10                          % Our default
\font\bf=cmbx10                         % Redefinition
                         % Redefinition
\font\it=cmti10                         % Redefinition
\font\trbig=cmbx10 scaled 1500          % Main Title
                          % Theorems                       
\font\tiny=cmr8                         % Running title
% --------------------------------------------------------------------
             % Math Sets eg R -> |R
                 % bold in math mode
                % small bold in math mode
\def\ng{>\kern -9pt|\kern 9pt}          % not greater than
                       % bra  <  math mode
                       % ket  >  math mode
\def\hi#1#2{$#1$\kern -2pt-#2}          % hyphen \hi{N}{body} = N-body
\def\hy#1#2{#1-\kern -2pt$#2$}          % hyphen hy{large}{N} = large-N

\def\half{{1 \over 2}}

% ----------------------------------------------------------------------

  %  QED
 % SQUARE 
% ---------------------------------------------------------------------- 
\output={\shipout\vbox{\makeheadline
                                      \ifnum\the\pageno>1 {\hrule}  \fi 
                                      {\pagebody}   
                                      \makefootline}
                   \advancepageno}

\headline{\noindent {\ifnum\the\pageno>1 
                                   {\tiny \ptitle\hfil page~\the\pageno}\fi}}
\footline{}
% ---------------------------------------------------------------------
\newcount\zz  \zz=0  % switch for printing references
\newcount\q   %  reference number
\newcount\qq    \qq=0  % starting reference number-1   (usually zero)

\def\pref #1#2#3#4#5{\frenchspacing \global \advance \q by 1     % paper reference
    \edef#1{\the\q}
       {\ifnum \zz=1 { %
         \item{[\the\q]} 
         {#2} {\bf #3},{ #4.}{~#5}\medskip} \fi}}

\def\bref #1#2#3#4#5{\frenchspacing \global \advance \q by 1     % book reference
    \edef#1{\the\q}
    {\ifnum \zz=1 { %
       \item{[\the\q]} 
       {#2}, {\it #3} {(#4).}{~#5}\medskip} \fi}}

\def\gref #1#2{\frenchspacing \global \advance \q by 1  % general reference
    \edef#1{\the\q}
    {\ifnum \zz=1 { %
       \item{[\the\q]} 
       {#2}\medskip} \fi}}

 \def\sref #1{~[#1]}

\def\references#1{\zz=#1
   \parskip=2pt plus 1pt   % default is 0pt plus 1pt       
   {\ifnum \zz=1 {\noindent \bf References \medskip} \fi} \q=\qq
\pref{\rosen}{M. E. Rose and R. R. Newton, Phys. Rev.}{82}{470 (1951)}{}
\pref{\ken}{P. Kennedy, J. Phys. A}{35}{689 (2002)}{}
\pref{\frnk}{J. Franklin, Mod. Phys. Lett. A}{14}{2409 (1999)}{}
\pref{\dfr}{A. S. De Castro, J. Franklin, Int. J. Mod. Phys. A}{15}{4355 (2000)}{}
\pref{\lmc}{L. Micu, Mod. Phys. Lett. A}{18}{2895 (2003)}{}
\pref{\bsk}{S. K. Bose, A. Schulze-Halberg, M. Singh, Phys. Lett. A}{287}{321 (2001)}{}
\pref{\bose}{S. K. Bose, A. Schulze-Halberg, Mod. Phys. Lett. A}{15}{1583 (2000)}{}
\pref{\giam}{J. Giammarco and J. Franklin, Phys. Rev. A}{36}{5839 (1987)}{}
\pref{\mosh}{M. Moshinsky and A. Sharma, J. Phys. A}{31}{397 (1998)}{}
\pref{\hall}{R. L. Hall, Phys. Rev. Letters}{83}{468 (1999)}{}
\pref{\gang}{G. Chen, Phys. Rev. A}{71}{024102 (2005)}{}
\pref{\cal}{A. Calageracos and N. Dombey, Phys. Rev. Lett.}{93}{180405 (2004)}{}
\pref{\dong}{S. H. Dong, X. W. Hou and Z. Q. Ma, Phys. Rev. A}{58}{2160 (1998)}{}
\pref{\lin}{Q. C. Lin, Phys. Rev. A}{57}{3478 (1998)}{}
\pref{\poli}{N. Poliatzky, Phys. Rev. Lett.}{70}{2507 (1993)}{}
\pref{\polit}{N. Poliatzky, Phys. Rev. Lett.}{76}{3655 (1996)}{}
\pref{\zma}{Z. Q. Ma, Phys. Rev. Lett.}{76}{3654 (1996)}{}
\pref{\guzo}{X. Y. Gu, Z. Q. Ma and S. H. Dong, Phys. Rev. A}{67}{062715 (2003)}{}
\pref{\hdong}{S. H. Dong and Z. Q. Ma, Phys. Lett. A}{312}{78 (2003)}{}
\pref{\dsng}{S. H. Dong and M. Lozada-Cossacu, Phys. Lett. A}{330}{168 (2004)}{}
\pref{\aldi}{A. D. Alhaidari, Phys. Rev. Lett.}{87}{210405 (2001)}{}
\pref{\laldi}{A. D. Alhaidari, Phys. Rev. Lett.}{88}{188901 (2002)}{}
\pref{\daldi}{A. D. Alhaidari, J. Phys. A}{34}{9827 (2001)}{}
\pref{\idaldi}{A. D. Alhaidari, J. Phys. A}{35}{6267 (2002)}{}
\pref{\jgou}{J. Y. Gou, X. Z. Fank and F. X. Xu, Phys. Rev. A}{66}{062105 (2002)}{}
\pref{\jogou}{J. Y. Gou, J. Meng and F. X. Xu, Chin. Phys. Lett.}{20}{602 (2003)}{}
\pref{\chen}{G. Chen, Phys. Lett. A}{328}{116 (2004)}{}
\pref{\dsgo}{S. H. Dong, J. Phys. A}{36}{4977 (2003)}{}
\pref{\sdong}{S. H. Dong, G. H. Sun and D. Popov, J. Math. Phys.}{44}{4467 (2003)}{}

\pref{\acka}{E. Ackad and M. Horbatsch, J. Phys. A.}{38}{3157 (2005)}{}

\pref{\ajo}{A. Joseph, Rev. Mod. Phys.}{39}{829 (1967)}{}
\pref{\yjg}{Y. Jiang, J. Phys. A.}{38}{1157 (2005)}{}
\pref{\gux}{X-Y. Gu et all, Phys. Rev. A}{67}{062715 (2003)}{}
\pref{\chs}{H. Ciftci, R. L. Hall and N. Saad, J. Phys. A: Math. Gen. }{36}{11807 (2003)}{}
\pref{\frn}{F. M. Fernandez, J. Phys. A: Math. Gen. }{37}{6173 (2004)}{}
\pref{\chss}{H. Ciftci, R. L. Hall and N. Saad, J. Phys. A: Math. Gen. }{38}{1147 (2005)}{}
\pref{\shc}{T. Barakat, K. Abodayeh, A. Mukheimer, J. Phys. A: Math. Gen. }{38}{1299 (2005)}{}
\pref{\halla}{R. L. Hall, J. Math. Phys.}{28}{457 (1987)}{}
\pref{\hallb}{R. L. Hall, J. Math. Anal. Apps.}{145}{365 (1990)}{}
\bref{\mes}{A. Messiah}{Quantum Mechanics II}{North Holland, Amsterdam, 1962}{The Dirac equation for central fields is given on page 928}
\pref{\empa}{C. H. Mehta and S. H. Patil, Phys. Rev. A}{17}{34 (1978)}{}

 }

 \references{0}    % Initialization of reference numbers
% ------------------------------------------------------------ end our ref.tex ---------------
% date stamp
% ----------------------------------------
%\ppl{\bf [ 12 May 2005 (dirhak.tex) ]}
% ------------------------------------------------------------------------- 
% preprint data using \ppl 
% ------------------------------------------------------------------------- 
\ppl{CUQM-109}
\ppl{math-ph/0505069} 
\ppl{May 2005}\medskip 
% ------------------------------------------------------------------------- 
\hskip 1cm
% ------------------------------------------------------ 
%   Title page and Abstract
% ------------------------------------------------------
\vskip 0.5 true in
\centerline{\bf\trbig\ptitle}
\bigskip
\centerline{Hakan Ciftci$^{1,3}$, Richard L. Hall$^{1}$ and Nasser Saad$^{2}$}
\medskip
\vskip 0.25 true in
\medskip
{\leftskip=0pt plus 1fil
\rightskip=0pt plus 1fil\parfillskip=0pt
\obeylines
$^{1}$Department of Mathematics and Statistics, Concordia University,
1455 de Maisonneuve Boulevard West, Montr\'eal,
Qu\'ebec, Canada H3G 1M8.\par}
\medskip
{\leftskip=0pt plus 1fil
\rightskip=0pt plus 1fil\parfillskip=0pt
\obeylines
$^{2}$Department of Mathematics and Statistics,
University of Prince Edward Island,
550 University Avenue, Charlottetown,
PEI, Canada C1A 4P3.\par}
\medskip
{\leftskip=0pt plus 1fil \rightskip=0pt plus 1fil\parfillskip=0pt
\obeylines $^{3}$Gazi Universitesi, Fen-Edebiyat Fak\"ultesi, Fizik
B\"ol\"um\"u, 06500 Teknikokullar, Ankara, Turkey.\par}

\vskip 0.5 true in
\baselineskip=18 true pt
% ---------------------------------------------------------------------
\centerline{\bf Abstract}\medskip
% ---------------------------------------------------------------------
We consider a single particle which is bound by a central potential and obeys the Dirac equation in $d$ dimensions. We first apply the asymptotic iteration method to recover the known exact solutions for the pure Coulomb case. For a screened-Coulomb potential and for a Coulomb plus linear potential with linear scalar confinement, the method is used to obtain accurate approximate solutions for both eigenvalues and wave functions.
\medskip\noindent PACS~~03.65.Pm,~31.15.Bs,~31.30.Jv.

\np
% ------------------------------------------------------ 
  \title{1.~~Introduction}
% ------------------------------------------------------
The Dirac equation plays a fundamental role in relativistic quantum mechanics. The equation can be solved exactly for a very few potentials. An early and very detailed analysis of the Dirac spectrum for central potentials has been given by Rose and Newton\sref{\rosen}. The solution for the pure Coulomb field is well-known. Exact solutions are known for some other specific cases, such as the Woods-Saxon potential\sref{\ken}. It is also possible to find classes of exact solutions under very special conditions\sref{\frnk-\bose}.  Unlike the corresponding Schr\"odinger operator, the Dirac Hamiltonian is not bounded below, and its spectrum can not be defined variationally. In spite of this it is still possible to find ways to use the variational idea, for example by a saddle-point analysis\sref{\giam} and by an analysis of the Schr\"odinger limit in a oscillator basis\sref{\mosh}. Some progress has been made in the establishment of comparison theorems for the Dirac equation without invoking a variational assumption at all\sref{\hall,\gang}, and this has allowed spectral envelope methods to be used; however, at present, such results apply only to node-free states.  Theorems of the Levinson type have now been proved for the Dirac equation in two, three, and $d$ dimensions\sref{\cal-\guzo}. Solutions of the Dirac equation with shape-invariant potentials have been found by a variety of exact and approximate methods\sref{\hdong-\sdong}.  Meanwhile, numerical methods for solving the Dirac equation are continually sought, such as a recent approach by a mapped Fourier grid method\sref{\acka}. Effective non-variational approximation methods therefore remain an important area for investigation.\medskip

In this paper we first consider the Dirac equation for central potentials in 
$d$ dimensions.  This problem was formulated some decades ago, by Joseph\sref{\ajo}. More recently, Yu Jiang\sref{\yjg}, for example, has studied the problem and has obtained a pair of radial equations similar to the well-known case of three dimensions; these equations can be solved exactly for the pure Coulomb field. The $d$-dimensional angular momentum problem for the Dirac equation has also been studied by group-theoretical methods\sref{\gux}.  We use these results in section~2. The main purpose of the present paper is to apply the asymptotic iteration method (AIM) to these central-field problems. AIM was first developed by the present authors\sref{\chs} for solving second order linear differential equations, including Schr\"odinger's. Later the method has been applied to a variety of problems\sref{\frn-\shc}.  In preparation for the Dirac application, we first extend the method in section~3 to treat {\it systems} of homogeneous linear differential equations.  In section~4 the case of the Dirac Coulomb problem in $d$ dimensions is then treated and solved exactly.  In sections~5 and 6 we study screened Coulomb problems, and also the linear plus Coulomb potential.  For the latter problem, the scalar linear part must dominate the vector linear part in order for discrete eigenvalues to exist\sref{\halla, \hallb}.
\np %%%%%%%%%%%%%%%%%% KLUDGE %%%%%%%%%%%%%%%%%%%%%%%%%%%%%%%%%%%%
% ------------------------------------------------------
  \title{2.~~The Dirac equation in $d$ dimensions}
% ------------------------------------------------------
The Dirac equation for a central field in $d$ dimensions can be written with $\hbar=c=1$ as
$$i{{\partial \Psi}\over{\partial t}} =H\Psi, H=\sum_{j=1}^{d}{\alpha_{j}p_{j}}+\beta (m+U(r))+V(r),\eqno{(2.1)}$$
where $m$ is the mass of the particle, $V(r)$ is a spherically symmetric vector potential, $U(r)$ is a spherically symmetric scalar potential, and $\{\alpha_{j}\}$ and $\beta$  are the usual Dirac matrices satisfying anti-commutation relations (the identity matrix is implied after the vector potential term $V(r)$). After some algebraic calculations (details can be found for example in\sref{\yjg}), one obtains the following first-order linear coupled differential equations 
$$
\eqalignno{
{{dG}\over{dr}}&=-{{k_{d}}\over{r}}G+(E+m-V(r)+U(r))F&(2.2)\cr
{{dF}\over{dr}}&=-(E-m-V(r)-U(r))G+{{k_{d}}\over{r}}F&(2.3)}$$
These are known as the radial Dirac equations in $d$ dimensions, where $k_{d}=\tau(j+{{d-2}\over{2}})$,  and $\tau = \pm 1.$  We note that the variable $\tau$ is sometimes written $\omega$, as, for example, in the book by Messiah\sref{\mes}.  As an example, we suppose that the particle is moving in a pure vector Coulomb field, that is to say, $V(r)=-{{A}\over{r}}$ and $U(r)=0$. In this case Eqs.(2.2) and (2.3) can be written as follows
$$
\eqalignno{
{{dG}\over{dr}}&= -{{k_{d}}\over{r}}G+\left(m+E+{{A}\over{r}}\right)F&(2.4)\cr
{{dF}\over{dr}}&= \left(m-E-{{A}\over{r}}\right)G+{{k_{d}}\over{r}}F.&(2.5)}$$
First of all, we have to obtain some asymptotic forms for $G(r)$ and $F(r)$ functions. At small $r$, $G(r)$ can be written as the following Euler equation
$${{d^{2}G}\over{dr^{}}}= -{{1}\over{r}}{{dG}\over{dr}}+\left({{k_{d}^{2}-A^{2}}\over{r^{2}}}\right)G.\eqno{(2.6)}$$
It is clear that the solution of Eq.(2.6) is: $G(r)=r^{\gamma}$, where $\gamma=\sqrt{k_{d}^{2}-A^{2}}$. When a similar analysis at small $r$ is made for $F(r)$, one finds the same general results. At large $r$, we obtain the following asymptotic differential equation for both $G(r)$ and $F(r)$:
${{d^{2}H(r)}\over{dr^{2}}}=(m^{2}-E^{2})H(r)$, where $H(r)$ is either $G(r)$ or $F(r).$ We conclude that $H(r)\sim\exp(-{\sqrt{m^{2}-E^{2}}r})$. We therefore adopt the following representations for these radial functions
$$
\eqalignno{
G(r)&=\sqrt{m+E}~r^{\gamma}\exp\left({-r\sqrt{m^{2}-E^{2}}}\right)(\phi_{1}+\phi_{2})&(2.7)\cr
F(r)&=\sqrt{m-E}~r^{\gamma}\exp\left({-r\sqrt{m^{2}-E^{2}}}\right)(\phi_{1}-\phi_{2})&(2.8)}$$
We now substitute these equations into the Eqs.(2.4) and (2.5) and use the notation $r=r_{1}\rho$ to find:
$$
\eqalignno{
{{d\phi_{1}}\over{d\rho}}&=\left(1-{{a+\gamma}\over{\rho}}\right)\phi_{1}-\left({{b+k_{d}}\over{\rho}}\right)\phi_{2}&(2.9)\cr
{{d\phi_{2}}\over{d\rho}}&=\left({{b-k_{d}}\over{\rho}}\right)\phi_{1}+\left({{a-\gamma}\over{\rho}}\right)\phi_{2},&(2.10)}$$
where $r_{1}={{1}\over{2\sqrt{m^{2}-E^{2}}}}$, $a=2EAr_{1}$ and $b=2mAr_{1}$. It is possible, of course, to solve these equations by using a power series method.  However, we shall solve the equations by means of AIM, which was developed originally for second order linear differential equations. In the next section, we extend the scope of AIM to apply to first order linear coupled differential equations generally; then in section~4 we apply the results obtained to the specific Dirac radial equations (2.9) and (2.10) above.
 
% ------------------------------------------------------
  \title{3.~~Asymptotic iteration method for first order linear coupled differential equations}
% ------------------------------------------------------
We consider the following first order linear coupled differential equations
$$
\eqalignno{
\phi^{\prime}_{1}&= \lambda_{0}(x)\phi_{1}+s_{0}(x)\phi_{2}&(3.1)\cr
\phi^{\prime}_{2}&= \omega_{0}(x)\phi_{1}+p_{0}(x)\phi_{2},&(3.2)}$$
where $\{\prime\}$ represents the derivative with respect to $x,$ and $\lambda_0(x)$, $s_0(x)$, $\omega_0(x),$ and $p_0(x)$ are sufficiently differentiable in appropriate domains. If we differentiate (3.1) and (3.2) with respect to $x$, we find that
$$
\eqalignno{
\phi^{\prime\prime}_{1}&= \lambda_{1}(x)\phi_{1}+s_{1}(x)\phi_{2}&(3.3)\cr
\phi^{\prime\prime}_{2}&= \omega_{1}(x)\phi_{1}+p_{1}(x)\phi_{2},&(3.4)}$$
where 

$$
\eqalign{
\lambda_{1}&=\lambda^{\prime}_{0}+\lambda^{2}_{0}+s_{0}\omega_{0},\cr 
s_{1}&=s^{\prime}_{0}+\lambda_{0}s_{0}+s_{0}p_{0},\cr
\omega_{1}&=\omega^{\prime}_{0}+\lambda_{0}\omega_{0}+p_{0}\omega_{0},\cr
p_{1}&=p^{\prime}_{0}+p^{2}_{0}+s_{0}\omega_{0}.
}$$
Similarly, if we calculate the $(n+2)^{th}$ derivative, $n=1,2,\dots$, we have
$$
\eqalignno{
\phi^{(n+2)}_{1}&= \lambda_{n+1}(x)\phi_{1}+s_{n+1}(x)\phi_{2}&(3.5)\cr
\phi^{(n+2)}_{2}&= \omega_{n+1}(x)\phi_{1}+p_{n+1}(x)\phi_{2},&(3.6)}$$
where 
$$
\eqalign{
\lambda_{n+1}=\lambda^{\prime}_{n}+\lambda_{n}\lambda_{0}+s_{n}\omega_{0},\cr 
s_{n+1}=s^{\prime}_{n}+\lambda_{n}s_{0}+s_{n}p_{0},\cr
\omega_{n+1}=\omega^{\prime}_{n}+\omega_{n}\lambda_{0}+p_{n}\omega_{0},\cr
p_{n+1}=p^{\prime}_{n}+\omega_{n}s_{0}+p_{n}p_{0}.
}$$
From the ratio of the $(n+2)^{th}$ and $(n+1)^{th}$ derivatives of $\phi_{1}$ we have
$${d\over dx}\ln\left(\phi_{1}^{(n+1)}\right)={\phi_{1}^{(n+2)}\over \phi_{1}^{(n+1)}}=
{\lambda_{n+1}(\phi_{1}+{s_{n+1}\over \lambda_{n+1}}\phi_{2})\over
\lambda_{n}(\phi_{1}+{s_{n}\over \lambda_{n}}\phi_{2})}.\eqno(3.7)$$
An exactly similar result can be obtained for $\phi_{2}$. However, to solve the system given in (3.1) and (3.2), one of these conditions is sufficient. We now introduce the `asymptotic' aspect of the method. If we have, for sufficiently large $n$,
$${s_{n+1}\over \lambda_{n+1}}={s_{n}\over \lambda_{n}} := \alpha,\quad n=1,2,3,\dots\eqno(3.8)$$
then (3.7) reduces to
${d\over dx}\ln(\phi_{1}^{(n+1)})={\lambda_{n+1}\over \lambda_{n}}$, which yields
$$\phi^{(n+1)}_{1}(x)=
C_1\exp\bigg(\int\limits^x{\lambda_{n+1}(t)\over
\lambda_{n}(t)}dt\bigg) = C_1\lambda_{n}\exp\left(\int\limits^x(\alpha\omega_{0}+\lambda_0)dt\right),\eqno(3.9)$$
where $C_1$ is the integration constant. After substituting Eq.(3.9) into $\phi^{(n+1)}_{1}= \lambda_{n}(x)\phi_{1}+s_{n}(x)\phi_{2}$, we get
$$\phi_{1}+\alpha(x)\phi_{2}=C_1\exp\left(\int\limits^x(\alpha\omega_{0}+\lambda_0)dt\right).\eqno{(3.10)}$$
Using Eqs.(3.10) and (3.2), we can obtain the general solution of $\phi_{2}(x)$ as follows
$$
\eqalignno{
\phi_{2}(x)&=\exp\left(\int\limits^{x}(p_{0}-\omega_{0}\alpha)dt\right)\left[C_2+ C_1\int\limits^{x}\left(\omega_{0}\exp\left(\int\limits^{t}(\lambda_{0}-p_{0}+2\omega_{0}\alpha)d\tau \right)dt\right)\right]&(3.11)}$$
Once we have obtained $\phi_{2}(x)$, it is easy to find $\phi_{1}(x)$ by using one of the coupled equations, or directly from Eq.(3.10). 
% -----------------------------------------------
\title{4.~~Solution of the Dirac Coulomb problem}
% -----------------------------------------------
\nl We now turn back to our first principal application. If we compare Eqs.(2.9) and (2.10) with Eqs.(3.1) and (3.2), we see that $\lambda_{0}(\rho)=1-{{a+\gamma}\over{\rho}}$, $s_{0}(\rho)=-{{b+k_{d}}\over{\rho}}$, $\omega_{0}(\rho)={{b-k_{d}}\over{\rho}}$ and $p_{0}(\rho)={{a-\gamma}\over{\rho}}$. By using our iteration formulas and the iteration termination condition given in Eq.(3.8), we find that
\nl $a=\gamma, 1+\gamma, 2+\gamma,...$ this means that $a=n+\gamma$, where $n=0, 1, 2, ...$. In this case $b$ satisfies the following relations
$$
\eqalignno{
n&=0;~a=\gamma;~b=-k_{d}&\cr
n&=1;~a=1+\gamma;~b=-k_{d},~\pm\sqrt{1+2\gamma+k_{d}^{2}}&\cr
n&=2;~a=2+\gamma;~b=-k_{d},~\pm\sqrt{3+2\gamma+k_{d}^{2}},~\pm\sqrt{4+4\gamma+k_{d}^{2}}&\cr
n&=3;~a=3+\gamma;~b=-k_{d},~\pm\sqrt{5+2\gamma+k_{d}^{2}},~\pm\sqrt{8+4\gamma+k_{d}^{2}},~\pm\sqrt{9+6\gamma+k_{d}^{2}}.&}$$
In general we have, for $a=n+\gamma,$ that $b=\pm\sqrt{k_{d}^{2}+s(2n-s)+2s\gamma}$, where $s=0, 1, 2,...n$. We know from Eqs.(2.9) and (2.10) that $a=2EAr_{1}$, $b=2mAr_{1},$ and $r_{1}={{1}\over{2\sqrt{m^{2}-E^{2}}}}$. When we use these equations, we find the following two different expressions for the energy
$$E=\pm {{m}\over{\sqrt{1+\left({{A}\over{n+\gamma}}\right)^{2}}}}\eqno(4.1)$$
and
$$E=\pm m\sqrt{1-{{A^{2}}\over{k_{d}^{2}+s(2n-s)+2s\gamma}}}.\eqno(4.2)$$
But these expressions must be equal. From their equality, $s$ must be $n$ or $n+2\gamma$. We know that both $n$ and $s$ are integers, but $\gamma$ is not an integer, so we find that $s=n$. In this case $b$ becomes $b=\pm\sqrt{k_{d}^{2}+n^{2}+2n\gamma}$, but when $n=0$, $b$ must be $-k_{d}$. Thus the energy has the following form
$$E=\pm m\left[1+\left({{A}\over{n+\sqrt{k^{2}_{d}-A^{2}}}}\right)^{2}\right]^{-{1/2}},\eqno(4.3)$$
where we use $\gamma=\sqrt{k^{2}_{d}-A^{2}}$ and $n=0,1,2,3,...$. If we define the principal quantum number as $n_{r}=n+|k_{d}|-{{d-3}\over{2}}=1, 2, 3,...$, we recover the following well-known formula for the Coulomb energy 
$$E=\pm m\left[1+\left({{A}\over{n_{r}-|k_{d}|+{{d-3}\over{2}}+\sqrt{k^{2}_{d}-A^{2}}}}\right)^{2}\right]^{-{1/2}}.\eqno(4.4)$$
% ------------------------------------------------------
  \title{5.~~Coulomb wave functions}
% ------------------------------------------------------
In this section we obtain the Coulomb eigenfunctions by using Eqs.(3.14) and (3.11). Eq.(3.11) includes two independent solutions for $\phi_{2}(x)$. The first factor of the expression in Eq.(3.11) generally represents the physical solution, so we use this factor as the wave-function generator for our model. If we rewrite this, we have
$$
\eqalignno{
\phi_{2}(x)&=C_{2}\exp\left(\int\limits^{x}(p_{0}-\omega_{0}\alpha)dt\right),&(5.1)}$$
where $C_{2}$ is an integration constant which can be determined by normalization. If we use our iteration procedure, we find the following results for $\phi_{2}$:
$$
\eqalignno{
\phi_{2}(\rho)&=1,~n=0\cr
\phi_{2}(\rho)&=-(2\gamma+1)(1-{{\rho}\over{2\gamma+1}}),~n=1\cr
\phi_{2}(\rho)&=(2\gamma+1)(2\gamma+2)(1-{{2\rho}\over{2\gamma+1}}+{{\rho^{2}}\over{(2\gamma+1)(2\gamma+2)}}),~n=2\cr
\phi_{2}(\rho)&=-(2\gamma+1)(2\gamma+2)(2\gamma+3)(1-{{3\rho}\over{2\gamma+1}}+{{3\rho^{2}}\over{(2\gamma+1)(2\gamma+2)}}-{{\rho^{3}}\over{(2\gamma+1)(2\gamma+2)(2\gamma+3)}}),~n=3.}$$
We see from these results that the general formula for $\phi_{2}(\rho)$ can be written as follows
$$
\eqalignno{
\phi_{2}(\rho)&=(-1)^{n}{{(2\gamma+n)!}\over{(2\gamma)!}}C_{2}~{_1F_{1}(-n,2\gamma+1,\rho)},&(5.2)}$$
where $_{1}F_{1}$ is the confluent hypergeometric function; since the first argument is a negative integer, the function is a polynomial of degree $n$. We can now calculate $\phi_{1}(\rho)$. For this task we use Eq.(3.10), but like $\phi_{2}(\rho)$, the solution of Eq.(3.10) has two parts, one of them is a polynomial and the other one is an infinite series: we choose the polynomial solution. Thus we find that $\phi_{1}(\rho)=-\alpha(\rho)\phi_{2}(\rho)$, where $\alpha(\rho)={{s_{n}}\over{\lambda_{n}}}$. If we calculate $\phi_{1}(\rho),$ using the above equation, we find the following results:
$$
\eqalignno{
\phi_{1}(\rho)&=0,~n=0\cr
\phi_{1}(\rho)&=(k_{d}+b),~n=1\cr
\phi_{1}(\rho)&=-(k_{d}+b)(2\gamma+1)(1-{{\rho}\over{2\gamma+1}}),~n=2\cr
\phi_{1}(\rho)&=(k_{d}+b)(2\gamma+1)(2\gamma+2)(1-{{2\rho}\over{2\gamma+1}}+{{\rho^{2}}\over{(2\gamma+1)(2\gamma+2)}}),~n=3.}$$
We conclude from these results that
$$
\eqalignno{
\phi_{1}(\rho)&=(-1)^{n+1}(b+k_{d}){{(2\gamma+n-1)!}\over{(2\gamma)!}}C_{2}~{_1F_{1}(1-n,2\gamma+1,\rho)},&(5.3)}$$
where $b={{mA}\over{\sqrt{m^{2}-E^{2}}}}$. After obtaining $\phi_{1}$ and $\phi_{2}$, we can recover the radial functions $G(r)$ and $F(r)$ in the well-known form by using Eqs.(2.7) and (2.8). Thus, we have the complete solution of the Dirac equation for the Coulomb problem. In the next sections, we will discuss the solution of the Dirac equation for a screened-Coulomb potential and Coulomb plus linear potential with a linear scalar confinement.
% ------------------------------------------------------
  \title{6.~~Dirac equation for a Screened-Coulomb potential in 3-dimensions}
% ------------------------------------------------------
In this section we will study a screened-Coulomb potential in 3-dimensions. We use the Mehta and Patil potential\sref{\empa} which is suitable for large atoms. This potential is defined by
$$V(r)=-{{v_{1}}\over{r}}+{{v_{2}\lambda}\over{1+\lambda r}},\eqno{6.1}$$
where $v_{1}=Z\alpha$, $v_{2}=(Z-1)\alpha$ and $\lambda=0.98\alpha Z^{1/3}$,
and $\alpha\approx 1/137.036$ is the fine-structure constant.  For this case, the radial Dirac equations read
$$
\eqalignno{
{{dG}\over{dr}}&= -{{k}\over{r}}G+\left(m+E+{{v_{1}}\over{r}}-W(r)\right)F&(6.2)\cr
{{dF}\over{dr}}&= \left(m-E-{{v_{1}}\over{r}}+W(r)\right)G+{{k}\over{r}}F,&(6.3)}$$
where $W(r)={{v_{2}\lambda}\over{1+\lambda r}}$. The asymptotic behaviours of $F(r)$ and $G(r)$ are the same as for the pure Coulombic potential. For this reason, $G(r)$ and $F(r)$ can be written as follows
$$
\eqalignno{
G(r)&=r^{\gamma}\exp\left(-r{\sqrt{m^{2}-E^{2}}}\right)(\phi_{1}+\phi_{2})&(6.4)\cr
F(r)&=r^{\gamma}\exp\left(-r{\sqrt{m^{2}-E^{2}}}\right)(\phi_{1}-\phi_{2}),&(6.5)}$$
where $\gamma=\sqrt{k^{2}-v^{2}_{1}}$. After substituting these forms into  into Eqs.(6.2) and (6.3), we have
$$
\eqalignno{
{{d\phi_{1}}\over{dr}}&=\left(m+\sigma-{{\gamma}\over{r}}\right)\phi_{1}-\left(E-W(r)+{{v_{1}+k}\over{r}}\right)\phi_{2}&(6.6)\cr
{{d\phi_{2}}\over{dr}}&=\left(E-W(r)+{{v_{1}-k}\over{\rho}}\right)\phi_{1}+\left(\sigma-m-{{\gamma}\over{r}}\right)\phi_{2},&(6.7)}$$
where $\sigma=\sqrt{m^{2}-E^{2}}$. Now we can use the AIM procedure to obtain the eigenvalues. In preparation for the iteration process it helps first to remove the square-root expression in $\sigma$.  We do this by the following subsitutions
$$\epsilon = \sqrt{{{m-E}\over{m+E}}},\quad \sigma = {{2m\epsilon}\over{1+\epsilon^2}},\quad E = m\left({{\epsilon^2 -1}\over{\epsilon^2 + 1}}\right).\eqno{(6.8)}$$
 If we use our iteration formulas and the iteration termination condition given in Eq.(3.8), we can construct the following equation, which corresponds to Eq.(3.8):
$$\delta(r,\epsilon)=\lambda_{n+1}s_{n}-s_{n+1}\lambda_{n}=0.\eqno(6.9)$$
If the problem is exactly solvable, then $\delta(r, \epsilon) = \delta(\epsilon)$ is independent of $r$ and its vanishing gives us the exact results, as with the  pure Coulomb problem discussed in section 4. In cases that are not exactly solvable in closed form,  $\delta(r, \epsilon)$ depends on both $r$ and on $\epsilon.$ We then solve the equation $\delta(r_0, \epsilon) = 0$ for a suitable fixed $r=r_{0}$ point, which choice affects the convergence rate of the iteration (this choice is discussed in more detail in the next section). For the problem at hand we chose only one value, $r_{0}=2,$ which fixed choice led to fast convergence in all cases. As the iteration number increases, the eigenvalue estimates become more accurate. The results for the ground state energies (i.e for $1s_{1/2}$) with $m=1$ and various atomic numbers $Z$ are presented in Table~1.  These agree with the results obtained earlier in Ref.\sref{\hall}. 
\np %%%%%%%%%%%%%%% KLUDGE new page %%%%%%%%%%%%%%%%%%%%%%
% ------------------------------------------------------
  \title{7.~~Dirac equation for the Coulomb plus linear potential with a linear scalar term}
% ------------------------------------------------------
As a further test of the method, we turn in this section to a problem quite different from that of atomic physics: we study the Dirac equation in 3-dimensions in the case that the vectorial part of the potential is Coulomb plus linear, and the scalar part is linear. We should like to point out that exactly similar calculations can be carried out in arbitrary dimensions $d$. For this problem the radial Dirac equation is written as follows
$$\left[\vec\alpha.\vec p+\beta(m+U(r))+V(r)\right]\psi=E\psi,\eqno{(7.1)}$$
where $V(r)=-{{A}\over{r}}+B_{1}r$ and $U(r)=B_{2}r$, $A$ is positive, and $B_{2}>B_{1}$. The correspoding radial Dirac equations become
$$
\eqalignno{
{{dG}\over{dr}}&= -{{k}\over{r}}G+\left(m+E+{{A}\over{r}}+(B_{2}-B_{1})r\right)F&(7.2)\cr
{{dF}\over{dr}}&= \left(m-E-{{A}\over{r}}+(B_{2}+B_{1})r\right)G+{{k}\over{r}}F,&(7.3)}$$
where $k=\tau(j+{{1}\over{2}})$ and $\tau = \pm 1.$ Before starting to calculate the energy eigenvalues, we have to consider the asymptotic behaviour of $F(r)$ and $G(r)$ at the boundaries. First we consider small $r$. Since the Coulomb potential dominates in this region, the asymptotic behaviour is the same as for the Hydrogenic problem: that is to say, $F(r)$ and $G(r)$ behave as $r^{\gamma}$, where $\gamma=\sqrt{k^{2}-A^{2}}$. At large $r$, we find that $H^{\prime\prime}(r)\sim \left(2(mB_{2}+EB_{1})r+(B^{2}_{2}-B_{1}^{2})r^{2}\right)H(r)$, where $H(r)$ is $G(r)$ or $F(r)$. Thus, $H(r)\sim \exp\left(-\alpha r-{{{1}\over{2}}\beta r^{2}}\right)$, where $\beta=\sqrt{B_{2}^{2}-B_{1}^{2}}$ and $\alpha={{mB_{2}+EB_{1}}\over{\beta}}$.  Thus we now write $G(r)$ and $F(r)$ in the following way
$$
\eqalignno{
G(r)&= r^{\gamma}\exp\left(-\alpha r-{{{1}\over{2}}\beta r^{2}}\right)(\phi_{1}+\phi_{2})&(7.4)\cr
F(r)&= r^{\gamma}\exp\left(-\alpha r-{{{1}\over{2}}\beta r^{2}}\right)(\phi_{1}-\phi_{2}).&(7.5)}$$
After substituting Eqs.(7.4) and (7.5) into (7.2) and (7.3), we find
$$
\eqalignno{
{{d\phi_{1}}\over{dr}}&= \left(m+\alpha+(\beta +B_{2})r-{{\gamma}\over{r}}\right)\phi_{1}-\left(E+{{A+k}\over{r}}-B_{1}r\right)\phi_{2}&(7.6)\cr
{{d\phi_{2}}\over{dr}}&= \left(E+{{A-k}\over{r}}-B_{1}r\right)\phi_{1}+\left(\alpha-m-{{\gamma}\over{r}}+(\beta -B_{2})r\right)\phi_{2}.&(7.7)}$$
When we compare these equations with (3.1) and (3.2) we see that $\lambda_{0}(r)=m+\alpha+(\eta +B_{2})r-{{\gamma}\over{r}}$, $s_{0}(r)=-E-{{A+k}\over{r}}+B_{1}r$, $\omega_{0}(r)=E+{{A-k}\over{r}}-B_{1}r$ and $p_{0}(r)=\alpha-m-{{\gamma}\over{r}}+(\eta -B_{2})r$. We now have to solve $\delta(E,r)=0$ at a suitable $r_{0}$ point. Thus $r_0$ is a parameter of the method. For the present problem, we have found that any $r_{0}$ satisfying $1 < r_0 < 3 $ is satisfactory in the sense that the iteration sequence converges rapidly.  For non-relativistic problems, we have found earlier that $r_0$ can be chosen as the peak of a simple scale-optimized trial function; a value of $r_0$ found in this way for the the ground state is also effective for the excited states.  However, since we don't have a suitable variational principle for the Dirac case, the `scale' as represented by $r_0,$ is chosen by the convergence criterion.  For the specific example discussed here, we adopted the fixed value $r_{0}=1.5$ and we obtained $E$ for $A=1/2$, $B_{2}=0.2,$ and $B_{1}=0.1$. The results are exhibited in Table~2 (the spectroscopic labelling is explained below).  When $B_{2}$ is much bigger than $B_{1},$ AIM gives us more accurate results for small iteration numbers, but here, we calculate the energy eigenvalues for rather close values of $B_{2}$ and $B_{1},$ in order to test the effectiveness of the method.\medskip
For the Dirac equation with central potentials $l$ is not a good quantum number. However, a spectroscopic description of the states is still possible if we adopt the following convention\sref{\hall}. We recall from section~2 that $\tau = \pm 1.$  Meanwhile, the lower index $\ell = 0,1,2,\dots$ of the spherical harmonic $Y_{\ell}^m$ appearing\sref{\mes} in the upper two components of the Dirac 4-spinor 
is related to $j$ by $\ell = j+\half\tau,$  and the parity of the state is given by the formula $P = (-1)^{j+\half\tau} = (-1)^{\ell}$. If the number $n = 1,2,3,\dots$ counts the eigenvalues for each given value of $k = \tau(j+\half),$  we may then define the `principle quantum number' for all central potentials as $n_{r}=n+\ell.$  The number $\ell$ can then be represented by the usual atomic symbol $\{s,p,d,f..\}$.  With this convention we label a state by $n_{r}D_{j}$, where $D=s,p,d,f..;$ In the non-relativistic large-$m$ limit, this notation agrees with the usual Schr\"odinger description. 
% --------------------------------------------------------------
\title{8. Wavefunctions for the linear plus Coulomb problem}
% --------------------------------------------------------------
It is possible to obtain approximate wavefunctions for the potential with Coulombic vectorial and scalar linear parts. We calculate the wavefunctions approximately by using AIM in the following way. As in section 4, we use the first part of Eq.(3.11) to generate the wavefunction. In this case, the function $\alpha(r)w_{0}(r)$ which appears in the wave function generator can not be integrated analytically at every iteration, so instead of doing this, we first expand this function near $r=0$ and then integrate the {\it representation} to give an approximation for $\phi_{2}(r).$ Once we have $\phi_{2}(r)$, it is straightforward to obtain $\phi_{1}(r)$ by using $\phi_{1}(r)=-\alpha(r)\phi_{2}(r)$. After obtaining $\phi_{1}(r)$ and $\phi_{2}(r)$, we can find $G(r)$ and $F(r)$ by using Eqs.(7.4) and (7.5). Below we give an example of these calculations. We choose the $3p_{3/2}$ state in Table~3 and find the following corresponding wave functions:
$$
\eqalignno{
G(r)&\approx\left(\sum_{k=0}^{15}{a_{k}r^{k}}\right)r^{0.866025}\exp{\left(-2.41965r-0.0866025r^{2}\right)}&(8.1)\cr
F(r)&\approx\left(\sum_{k=0}^{15}{b_{k}r^{k}}\right)r^{0.866025}\exp{\left(-2.41965r-0.0866025r^{2}\right)},&(8.2)}$$
where $a_{k}$ and $b_{k}$ coefficients are given in Table~2. We tabulate more coefficients than are needed, in order to demonstrate the stability of the method. In Figure~1 we show Cartesian plots of the radial functions $G(r)$ and $F(r),$ and in Figure~2 we exhibit the Dirac spinor orbit\sref{\halla} defined by ($G(r)$, $F(r)$), $r\geq 0.$ The availability of the wave-function approximations makes these tasks straightforward.\medskip
%------------------------------------------------------------------------------
\title{9. Conclusion}
In this paper we have shown how AIM can be used to solve systems of two first order linear differential equations.  In cases where the system represents an eigenvalue problem, the method yields the eigenfunctions and the eigenvalues. If the exact wave function may be factored in the form of an asymptotic wave function multiplied by a polynomial, the problem can be solved exactly. In other cases, an approximate solution is found by forcing the vanishing of a certain function $\delta(r, E)$ after a finite number of iterations at a fixed expansion point $r = r_0.$  The range of values of $r_0$ which all lead to fast convergence is not narrow: for the problems discussed in this paper the range $1 < r_0 < 3$ was satisfactory; for highly excited states with Coulomb-like potentials, we have found larger values to be better.\medskip

In this paper we report applications of the method to bound states of the Dirac equation.  First of all, as a test, the known exact solutions of the Coulomb problem in $d$ dimensions were recovered.  The method was then applied to find the spectrum and wave functions for a screened-Coulomb potential, and also for a very different problem, namely a linear plus Coulomb potential with a scalar linear confining term. In all cases the method yielded fast convergence to accurate solutions.    
\medskip 
% ------------------------------------------------------   
   \title{Acknowledgments}
% ------------------------------------------------------
  \medskip
\noindent Partial financial support of this work under Grant Nos. GP3438 and GP249507 from the Natural Sciences and Engineering Research Council of Canada is gratefully acknowledged by two of us (respectively [RLH] and [NS]).
% ------------------------------------------------------ 
\np
\references{1}
% ------------------------------------------------------
\np
% ------------------------------------------------------
%  Table
% ------------------------------------------------------
\noindent {\bf Table 1}~~Ground state eigenvalues $E$ for the state $k=-1,~j={1/2}$ (with spectral description $1s_{\half}$) for the screened Coulomb potential. The energies $(E-1)m_e$, where $m_{e}=511.004$~KeV, are shown, along with corresponding accurate numerical values for comparison.  
\vskip 0.5 true in 

\def\vr{\vrule height 18 true pt depth 11 true pt}
\def\vra{\vr\hfill} \def\vrb{\hfill &\vra} \def\vrc{\hfill & \vr\cr\hrule}
\def\vrq{\vr\quad} 

$$\vbox{\offinterlineskip
 \hrule
\settabs
\+ \vrq \kern 0.4true in &\vrq \kern 0.7true in &\vrq \kern 0.9true in &\vrq \kern 0.7true in &\vrq \kern 0.7true in &\vr\cr\hrule

\+ \vra $Z$ \vrb $E$\vrb $(E-1)m_{e}$\vrb Numerical\vrc
\+ \vra 20 \vrb 0.991560\vrb -4.3129\vrb\ -4.3157\vrc 
\+ \vra 30 \vrb 0.979852\vrb -10.2957\vrb -10.2960\vrc
\+ \vra 40 \vrb 0.962675\vrb -19.0732\vrb -19.0732\vrc
\+ \vra 50 \vrb 0.939619\vrb -30.8549\vrb -30.8543\vrc
\+ \vra 60 \vrb 0.910139\vrb -45.9193\vrb -45.9189\vrc
\+ \vra 70 \vrb 0.873475\vrb -64.6548\vrb -64.6545\vrc
\+ \vra 80 \vrb 0.828543\vrb -87.6152\vrb -87.6148\vrc

}$$

\np
% ------------------------------------------------------
%  Table 2
% ------------------------------------------------------
\noindent {\bf Table~2}~~Eigenvalues $E$ of the Dirac Hamiltonian for the linear plus Coulomb vector potential $V(r) = -A/r + B_1 r$ and a linear scalar potential $U(r) = B_2~r$ in three dimensions, where $A=0.5,~B_{2}=0.2,~B_{1}=0.1$ The results are given in dimensionless units corresponding to $m = 1.$~ Accurate numerical results $E_{\rm num}$ are shown for comparison: this accuracy was obtained with 20 iterations.
\vskip 0.5 true in 

\def\vr{\vrule height 18 true pt depth 11 true pt}
\def\vra{\vr\hfill} \def\vrb{\hfill &\vra} \def\vrc{\hfill & \vr\cr\hrule}
\def\vrq{\vr\quad} 

$$\vbox{\offinterlineskip
 \hrule
\settabs
\+ \vrq \kern 0.4true in &\vrq \kern 0.7true in &\vrq \kern 0.7true in &\vrq \kern 0.7true in &\vrq \kern 0.7true in &\vr\cr\hrule

\+ \vra $k$ \vrb $n$\vrb States\vrb $E$\vrb $E_{\rm num}$\vrc
\+ \vra -1 \vrb 0\vrb $1s_{1/2}$\vrb 1.25819\vrb 1.25819\vrc 
\+ \vra -1 \vrb 1\vrb $2s_{1/2}$\vrb 1.87575\vrb 1.87575\vrc
\+ \vra -1 \vrb 2\vrb $3s_{1/2}$\vrb 2.29722\vrb 2.29722\vrc
\+ \vra 1 \vrb 0\vrb $1p_{1/2}$\vrb 1.70367\vrb 1.70367\vrc
\+ \vra 1 \vrb 1\vrb $2p_{1/2}$\vrb 2.15272\vrb 2.15272\vrc
\+ \vra 1 \vrb 2\vrb $3p_{1/2}$\vrb 2.51020\vrb 2.51029\vrc
\+ \vra -2 \vrb 0\vrb $2p_{3/2}$\vrb 1.74683\vrb 1.74683\vrc
\+ \vra -2 \vrb 1\vrb $3p_{3/2}$\vrb 2.19096\vrb 2.19096\vrc
\+ \vra -2 \vrb 2\vrb $4p_{3/2}$\vrb 2.54480\vrb 2.54486\vrc
\+ \vra 2 \vrb 0\vrb $2d_{3/2}$\vrb 2.03889\vrb 2.03889\vrc
\+ \vra 2 \vrb 1\vrb $3d_{3/2}$\vrb 2.41193\vrb 2.41193\vrc
\+ \vra 2 \vrb 2\vrb $4d_{3/2}$\vrb 2.72766\vrb 2.72762\vrc
\+ \vra -3 \vrb 0\vrb $3d_{5/2}$\vrb 2.04506\vrb 2.04506\vrc
\+ \vra -3 \vrb 1\vrb $4d_{5/2}$\vrb 2.42019\vrb 2.42019\vrc
\+ \vra -3 \vrb 2\vrb $5d_{5/2}$\vrb 2.73666\vrb 2.73665\vrc

}$$

\noindent {\bf Table 3}~~The Coefficients $a_{k}$ and $b_{k}$ in Eqs(8.1) and (8.2)
\vskip 0.5 true in 

\def\vr{\vrule height 18 true pt depth 11 true pt}
\def\vra{\vr\hfill} \def\vrb{\hfill &\vra} \def\vrc{\hfill & \vr\cr\hrule}
\def\vrq{\vr\quad} 

$$\vbox{\offinterlineskip
 \hrule
\settabs
\+ \vrq \kern 0.4true in &\vrq \kern 1true in &\vrq \kern 1true in &\vrq \kern 0.7true in &\vrq \kern 0.7true in &\vr\cr\hrule

\+ \vra $k$ \vrb $a_{k}$\vrb $b_{k}$\vrc
\+ \vra 0\vrb 1.7746\vrb -0.22540\vrc 
\+ \vra 1\vrb 3.34842\vrb -0.87777\vrc
\+ \vra 2\vrb 2.58401\vrb-1.16405\vrc
\+ \vra 3\vrb 0.89712\vrb -0.79148\vrc
\+ \vra 4\vrb $-4.42384.10^{-2}$\vrb -0.30087\vrc
\+ \vra 5\vrb -0.20572\vrb $-4.92821.10^{-2}$\vrc
\+ \vra 6\vrb -0.11569\vrb $1.18971.10^{-2}$\vrc
\+ \vra 7\vrb $-3.99651.10^{-2}$\vrb $1.06347.10^{-2}$\vrc
\+ \vra 8\vrb $-1.00424.10^{-2}$\vrb $3.82882.10^{-3}$\vrc
\+ \vra 9\vrb $-1.95538.10^{-3}$\vrb $9.31883.10^{-4}$\vrc
\+ \vra 10\vrb $-3.04816.10^{-4}$\vrb $1.71231.10^{-4}$\vrc
\+ \vra 11\vrb $-3.87895.10^{-5}$\vrb $2.48733.10^{-5}$\vrc
\+ \vra 12\vrb $-4.07966.10^{-6}$\vrb $2.92628.10^{-6}$\vrc
\+ \vra 13\vrb $-3.57403.10^{-7}$\vrb $2.82781.10^{-7}$\vrc
\+ \vra 14\vrb $-2.61991.10^{-8}$\vrb $2.26316.10^{-8}$\vrc
\+ \vra 15\vrb $-1.61008.10^{-9}$\vrb $1.50639.10^{-9}$\vrc

}$$

\np
\medskip

\hbox{\vbox{\psfig{figure=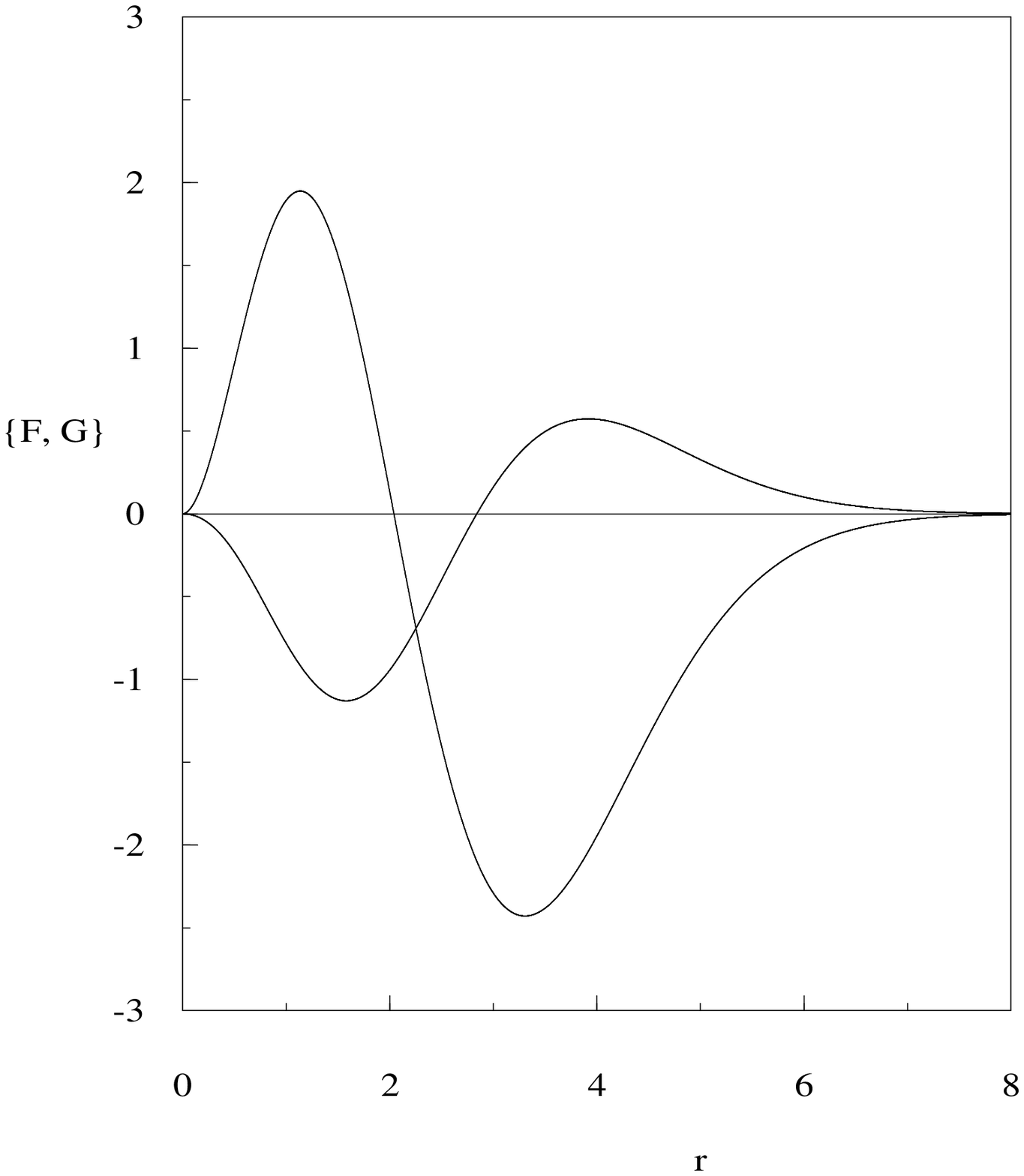,height=5in,width=5in,silent=}}}
\nl{\bf Figure~1.}~~Dirac radial functions $G(r)$ and $F(r)$,~$r > 0,$ in dimensionless units for 
$V(r) = -{{1}\over{2r}}+0.1r,$ $U(r) =0.2r,$ $k =-2,~ j = {{3}\over{2}},~ n = 2.$  This state can be described by the spectroscopic convention as $3p_{3/2}.$
\np
\hbox{\vbox{\psfig{figure=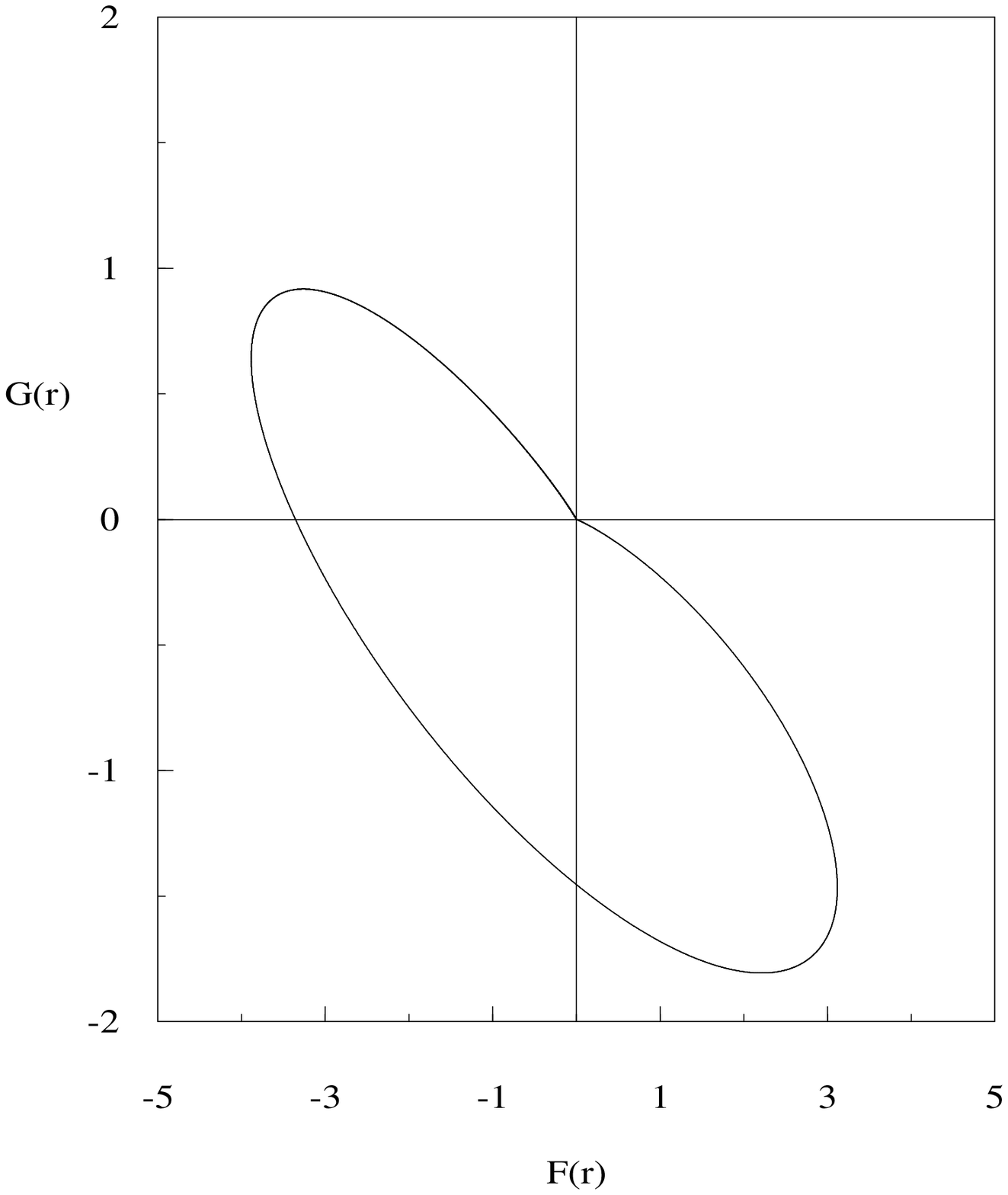,height=5in,width=5in,silent=}}}
\nl{\bf Figure 2.}~~Dirac spinor orbit $(F(r), G(r)),~r > 0,$  
in dimensionless units for the same example as in Figure~1

% -------------------------------------------------------------------------
\hfil\vfil
\end